\documentclass[aps,preprint]{revtex4}%
\usepackage{amsfonts}
\usepackage{amsmath}
\usepackage{amssymb}
\usepackage{graphicx}
\usepackage{pdfpages}
\usepackage{cleveref}
\usepackage{subcaption}%
\providecommand{\U}[1]{\protect\rule{.1in}{.1in}}
\providecommand{\U}[1]{\protect\rule{.1in}{.1in}}

\begin{document}
\preprint{HEP/123-qed}
\title{Holographic Dissipative Properties of Non-relativistic Black Branes
with Hyperscaling Violation}
\author{Huriye G\"{u}rsel}
\affiliation{}
\author{Mert Mangut}
\affiliation{}
\author{\.{I}zzet Sakall{\i}}
\affiliation{Physics Department, Arts and Sciences Faculty, Eastern Mediterranean
University, Famagusta, North Cyprus via Mersin 10, Turkey.}
\author{}
\affiliation{}

\begin{abstract}
In this work, we consider a class of hyperscaling violating Lifshitz-like black branes with metric scaling components $z=2$ and $\theta=-1$ whose corresponding holographic model can be treated as a non-relativistic fluid exhibiting Lifshitz-type symmetry. Having performed analytical calculations via the Klein-Gordon equation and the linear response theory, the experimental realizations of the concerned model, namely the transport coefficients, are found to behave as  $\eta \propto T^{3/2}$, $\sigma_{DC} \propto T^{3/2}$, and $\rho \propto T^{-3/2}$. The associated metric scaling exponents from the bulk theory are encrypted in the transport coefficients obtained for the holographic dual model. We believe that our analytical results can contribute to the endeavours in accomplishing a full understanding on the strongly coupled phenomena occurring in systems such as high temperature superconductors, the hypothetical magnetic monopoles, and liquid crystals.

\end{abstract}
\volumeyear{ }
\eid{ }
\date{\today}
\received{}

\maketitle
\tableofcontents

\section{Introduction}

One of the compelling problems which seems to sustain its complexity until
today is the complete physical explanation of the mechanism behind the
atomic nuclei. In 1931, Dirac stated that having a complete theory for
understanding the nature of atomic nuclei would be a rather challenging task
and would require a drastic revision of our fundamental understanding of
nature. Furthermore, he also added that constructing a theory directly from
the experimental data would go beyond the intelligence of humanity; and
thus, suggested the theorists of the future to look for indirect ways of
approaching the problem \cite{Dirac}. Today, a detailed study on various gravitational
models can enable one to achieve a deeper understanding of the systems in
nature where strong coupling exist, which supports Dirac's perspective.
Although quantum chromodynamics (QCD) is an effective and experimentally
consistent theory for explaining the high energy behavior of strongly
coupled structures, it confronts several challenges once the low energy
regime is of concern \cite{Maldacena,Witten}. AdS/CFT correspondence \cite{AdS}, which is a very well-established realization of the holographic principle \cite{Hoft,Susskind}, can be thought as
a channel to tackle rather complicated problems in one field with the tools
of the other. The holographic principle makes it possible to perceive the
entire physical phenomena occurring in nature as a whole and treats
seemingly-different models as the dual of each other. The low energy
behavior of strongly coupled systems can be considered as being equivalent
to the hydrodynamical properties of the event horizons of astronomical objects
which are commonly referred to as the membrane paradigm \cite{membrane1,membrane2}. With this
approach, one can aim for obtaining information regarding the low energy
behavior of strongly coupled systems via mapping the results of bulk
gravitational models on to the field theory of concern. \\

Quark confinement and chiral symmetry breaking are co-existing concepts in
low energy QCD, which are still in need of a consistent theoretical
explanation. There are numerous studies such as Refs. \cite{numeric1,numeric2,numeric3,numeric4,numeric5} where these phenomena are explained
via numerical simulations; however, there seems to be a
gap in literature for exact analytical approaches, as the usual perturbative
methods are not applicable in this regime. Throughout this work,
we investigate the electrically charged hyperscaling violating (\textit{hv})
Lifshitz-like black branes (BBs) in $(3+1)$-dimensional spacetime and find the
transport coefficients of the dual model, namely the shear viscosity $\eta$, DC-
conductivity $\sigma_{DC}$, and DC-resistivity $\rho_{DC}$, via exact
analytical methods. The bulk gravitational model of our concern consists of
a Maxwell field, a dilatonic scalar, a negative cosmological constant, and $%
N SU(2)$ Yang-Mills fields. For studies regarding \textit{hv}-spacetime structures, please refer to \cite{h1,h2,h3,h4,h5,h6,h7,h8,h9,h10,h11,h12}. In fluid/gravity correspondence, the dynamic critical exponent $z$ and the \textit{hv}-factor $\theta$ play a vital role in both characterizing the properties of the bulk model and determining the scaling behavior of the observables in the dual scenario. On the gravitational side, besides being subject to an overall \textit{hv}-factor, the metric also encounters a temporal anisotropy due to quantum critical phenomena. Such Lifshitz-like spacetimes correspond to the dual models, which experience continuous phase transitions \cite{exponent}.\\

In 1974, 't Hooft made a proposition that for any gauge theory, there exists a dual
string model in the large $N$ limit \cite{Hooft}. Thus, the Lifshitz-like BB of our
concern may be linked not only to the $(2+1)$-dimensional field theory model
that it corresponds to, but also to the associated one-spatial dimensional
dual string theory. At this point, we shall emphasize that both the
observational data and the theoretical framework of the Veneziano model \cite{Veneziano} indicate the likely possibility of having an underlying string structure to
hadronic matter \cite{dual2}. Once these relations are
investigated, one cannot go without noticing the relevance of the theory of
magnetic monopoles and the hadronic matter. In 1974, Mandelstam \cite{Mandelstam} proposed a
model in which the experimentally required confinement condition was
satisfied. During his work, he combined the Nielsen-Olesen interpretation of
ST \cite{NOST} and Nambu's idea \cite{Nambuu} of treating quarks as the carriers of magnetic charges.
Since then, there have been a great number of studies aiming to find out
ways for observing magnetic monopoles. Dirac put
forward the idea that the fundamental dissimilarity between electricity and
magnetism most likely arises from the same conceptual reasoning that causes
the electrons and protons to possess different properties \cite{Dirac}.
The non-Abelian massless monopoles in low energy effective
action of supersymmetric theories are thought to play a significant role in
problems existing in low energy behavior of QCD. In addition, it is
worth noting that the ground state of QCD can be treated as a dual
superconductor \cite{dual3,Mandelstam,dual2,dual}. Thus, based on the aforementioned discussions, we
suggest that it is highly likely for the analytically obtained transport
coefficients of the BB structure of our concern to carry information
regarding the unsolved problems in the theories of magnetic monopoles,
superconductors, and low energy behavior of QCD, in accordance with the
holographic principle. The dynamic scaling exponent of the model is chosen
to be $z=2$ so as to support superconducting fluctuations \cite{z}. Furthermore, theories with $z=2$ scaling describe multicritical points in certain liquid crystals and have been shown to arise at quantum critical points in toy models of the cuprate superconductors \cite{8}. On the other hand, the spatial dimension of our bulk spacetime is adopted as $d=3$. The reason behind this specific choice is to shift the perception of a three-spatial dimensional reality created by our minds (as a direct consequence of observing the macroscopic world) to a (2+1)-dimensional holographic scenario in which the two cases exhibit common properties. In Ref. \cite{Hooft2}, 't Hooft claims that to be able to construct a consistent quantum gravity model, the observable degrees of freedom should be described as if they were Boolean variables defined on a two-dimensional lattice, which also coincides with our specific choice of dimensionality. Consequently, the \textit{hv}-factor is determined by the specific $z$ and $D$ values of our choice, where $D$ stands for the overall dimensionality; i.e. $D=d+1$. \\

Our paper is organized as follows. In Section II, we briefly describe the properties of the (3+1)-dimensional gravitational model of our interest, with metric exponents $z=2$ and $\theta=-1$, which corresponds to the non-relativistic systems. Furthermore, we provide an exact solution for the associated Klein-Gordon equation (KGE) following Ref.\cite{biz}. For the absorption cross-section, decay rate, and greybody factor of the concerned model, the reader is referred to Ref.\cite{biz}. Section III, on the other hand, is devoted to the analytical evaluation of the dissipative properties of the dual model living on the boundary, which is the main focusing point of this study. To this end, the asymptotic behavior of KGE is used to evaluate the Green's function. Ultimately, the two scenarios are mapped to each other via the linear response theory; and in turn, the effect of the metric exponents on the observables of the dual theory is investigated. The final section is reserved for the conclusive remarks and for our future work plan.    

\section{Bulk Geometry}
In this section, we briefly review the bulk geometry of the holographic model of our concern, namely the non-Abelian charged Lifshitz-like BB with hyperscaling violation. These BBs arise as solutions to the theory possessing the
Lagrangian that goes as \cite{SourceBecar}%
\begin{equation}
\begin{aligned} \mathcal{L}=\sqrt{-g}[ R-\Lambda e^{-\lambda\phi}
&-\frac{1}{2}\left( \partial \phi\right)
^{2}-\overset{N}{\underset{k=1}{{\displaystyle\sum}}}\frac{1}{4g_{k}^{2}}e^{\lambda\phi}F_{k}^{2}\\
&-\frac{1}{4}e^{\lambda\phi
}\mathcal{F}]. \end{aligned}  \label{E3}
\end{equation}%

As can be seen from Eq. (\ref{E3}), the gravitational model described by the Lagrangian above consists of Einstein gravity, a cosmological constant $\Lambda$, a dilatonic scalar field $\phi$, a Maxwell field $\mathcal{A}$ whose associated field strength reads $\mathcal{F}=d\mathcal{A}$, and $N$ $SU(2)$ Yang-Mills field
$A_{k}^{a}$ with strength $F_{\mu \nu }^{a}$. Note that $F_{\mu \nu }^{a}=\partial _{\mu }A_{\upsilon}^{a}-\partial _{\upsilon }A_{\mu }^{a}+\epsilon ^{abc}A_{\mu
}^{b}A_{\upsilon }^{c}$, in which $a$ runs from 1 to $N$.\\ 

A class of \textit{hv} solutions with general dynamical exponent to the theory described by Lagrangian (\ref{E3}) can be written as
\begin{equation}
ds^{2}=r^{\theta }\left( -r^{2z}f(r)dt^{2}+\frac{dr^{2}}{r^{2}f(r)}+r^{2}%
\overset{D-2}{\underset{i=1}{\sum }}dx_{i}^{2}\right) ,  \label{E4}
\end{equation}%
where 
\begin{equation}
f(r)=1-\frac{q^{2}}{2\left( z-1\right) r^{2\left( z-1\right) }},  \label{E5}
\end{equation}
and
\begin{equation}
\theta =\frac{2}{D-2}[z-(D-1)].  \label{E5y}
\end{equation}
Note that $q$ represents the charge parameter determining the exact electric charge of the BB via $Q=\frac{\omega}{16\pi}q$. For the cases when $\theta \neq 0$, the system of concern breaks scale invariance to covariance which causes a power law scaling of thermodynamic
observables, relative to that of a conformal field theory
\cite{yeni,SCH}. The \textit{hv}-factor arises due to the presence of the dilaton which is assumed to have the ansatz
\begin{equation}
\phi=\frac{\theta}{\lambda}\log{r},
\end{equation}
with $\lambda$ denoting the 't Hooft coupling. On the other hand, the Maxwell field is taken to be in the form
\begin{equation}
\mathcal{A}=(\varphi_{0}+qr)dt,   
\end{equation}
where $\varphi_0$ can be treated as the gauge parameter. For the specific model of our interest, we choose the dynamical exponent and the
spacetime dimension as $z=2$ and $D=4$ respectively, which in turn results in $\theta =-1$. Thus,
metric (\ref{E4}) can be rearranged as 
\begin{equation}
ds^{2}=-N(r)dt^{2}+\frac{dr^{2}}{N(r)}+r\overset{2}{\underset{i=1}{\sum }}%
dx_{i}^{2},  \label{E5X}
\end{equation}%
whose metric function reads $N(r)=r^{3}f(r)$. Consequently, Eq.(\ref{E5}) reduces to $f(r)=1-\frac{q^{2}}{2r^{2}}
$. One shall record that metric (\ref{E5X}) is non-relativistic; and thus, the tools of AdS/CFT correspondence can be extended to obtain information about the observables of the dual non-relativistic CFT model. For the above $4D$ non-Abelian charged Lifshitz BB (\ref{E5X}), one can
compute the surface gravity \cite{wald} as 
\begin{equation}
\kappa_s =\left. \frac{1}{2}\frac{dN(r)}{dr}\right\vert _{r=r_{H}}=r_{H}^{2},
\label{E5XX}
\end{equation}%
in which the event horizon is $r_{H}=\pm q/\sqrt{2}$. Therefore, the corresponding Hawking
temperature \cite{hawr} results in
\begin{equation}
T=\frac{\kappa_s }{2\pi }=\frac{r_{H}^{2}}{2\pi }=\frac{q^{2}}{4\pi }\text{.}\label{9}
\end{equation}

Considering massless scalar perturbations of this bulk model, the associated KGE can be presented as
\begin{equation}
\frac{1}{\sqrt{-g}}\partial _{\mu }\left( \sqrt{-g}g^{\mu \upsilon }\partial
_{\nu }\varphi \right) =0,  \label{KG}
\end{equation}%
where $\varphi $ denotes the spin-$0$ field. The radial part of Eq.(\ref{KG}), which has already been evaluated in Ref. \cite{biz}, goes as
\begin{equation}
N(r)\frac{d^{2}R(r)}{dr^{2}}+(4r^{2}-2r_{H}^{2})\frac{dR(r)}{dr}+\left( \frac{%
\omega ^{2}}{N(r)}-\frac{\kappa ^{2}}{r}\right) R(r)=0,  \label{radial1}
\end{equation}%
where $-\kappa ^{2}$ is the eigenvalue of the Laplacian in the flat
base sub-manifold \cite{Becar} and $R(r)$ is the radial part of the scalar field based on the ansatz $\varphi (t,r,\vec{x})=\Phi(r)e^{i\vec{\kappa}\cdot \vec{x}}e^{-i\omega
t}$. In order to be able to solve Eq. (\ref{radial1}), one may change the variable into $\Tilde{z}=r^{-2}(r^{2}-r_{H}^{2})$ and set 
\begin{equation}
\Phi(\Tilde{z})=\Tilde{z}^{\alpha }(1-\Tilde{z})^{3/2 }G(\Tilde{z}).
\label{trans}
\end{equation}
This results in
\begin{equation}
\begin{aligned}
G(\Tilde{z})=C_{1}{}_{2}&F_{1}\left( a,b;c;\Tilde{z}\right)+
C_{2}\Tilde{z}^{1-c}\\
&{}_{2}F_{1}\left(
a-c+1,b-c+1;2-c;\Tilde{z}\right) ,  \label{iz1}    
\end{aligned}
\end{equation}%
with the relevant constants

\begin{equation*}
\alpha =-(i\omega /2\kappa _{s}),
\end{equation*}
\begin{equation}
a=\frac{5}{4}-\frac{i}{2\kappa _{s}}\left( \omega +\widehat{\omega }\right) ,
\label{e}
\end{equation}%
\begin{equation}
b=\frac{5}{4}-\frac{i}{2\kappa _{s}}\left( \omega -\widehat{\omega }\right) ,
\label{f}
\end{equation}%
where%
\begin{equation}
\widehat{\omega }=\sqrt{\omega ^{2}+\kappa _{s}\left( \kappa ^{2}-\frac{%
\kappa _{s}}{4}\right) }.  \label{o}
\end{equation}
For further details and the thermal radiation parameters of this model, please refer to Ref. \cite{biz}.

\section{Holographic Approach: Transport Coefficients of the Dual Model}
Having briefly mentioned the characteristic properties of the $4D$ semi-classical BB of our concern, we can now shift our focus towards the associated holographic dual model living on the boundary. At this point, one shall recall that the low energy behavior of strong interactions in the dual field theory are governed by the laws of hydrodynamics and their experimental realizations such as $\eta$, $\sigma_{DC}$, and $\rho_{DC}$ can be evaluated via linear response theory \cite{kaynak}. Throughout this section, we shall be using the tools of fluid/gravity correspondence to build a bridge between the gravitational model of our choice and its corresponding dual model; and consequently, the dissipative properties of the holographic dual model will be investigated via analytical methods.

\subsection{Shear Viscosity}
As a direct consequence of the fluid/gravity correspondence, the shear viscosity of the dual model of \textit{hv} Lifshitz-like BB can be found by using Kubo's formula \cite{Kubo,Czajka:2017bod,Moghaddas:2017itv}, which is given by
\begin{equation}
\eta=-\lim_{\omega\to 0}\frac{1}{\omega} Im \left [ G_{O_{+}}(\omega ,0)
\right ],  \label{E6}
\end{equation}
where $G_{O_{+}}$ denotes the retarded Green's function that can be obtained via investigating the asymptotic behavior of the solution of radial KGE (\ref{radial1}). For $r\rightarrow \infty $, Eq. (\ref%
{radial1}) reduces to 
\begin{equation}
\frac{d^{2}R}{dr^{2}}+\frac{4}{r}\frac{dR}{dr}=0,  \label{radial3}
\end{equation}
which allows us to express the radial solution at spatial infinity as 
\begin{equation}
R(r)=D_{1}+\frac{D_{2}}{r^{3}}, \label{r33}
\end{equation}
where $D_{1}=A_{2}C_{1}$ and 
$D_{2}=A_{1}C_{1}r_{H}^{3}$ with
\begin{equation}
A_{1}=\frac{\Gamma (c)\Gamma (c-a-b)}{\Gamma (c-a)\Gamma (c-b)},  \label{bir}
\end{equation}
and
\begin{equation}
A_{2}=\frac{\Gamma (c)\Gamma (a+b-c)}{\Gamma (a)\Gamma (b)}.  \label{iki}
\end{equation}
Then, by using $G_{O_{+}}=D_2/D_1$ \cite{H.Lu}, the two-point correlation function can be expressed in the form
\begin{equation}
G_{O_{+}}(\omega ,0)=\frac{8}{3}\frac{\Gamma \left ( \frac{5}{4}-i\tilde{Y}
\right)}{\Gamma \left ( -\frac{1}{4}-i\tilde{Y} \right)}\frac {\Gamma \left
( \frac{5}{4}-i\tilde{X} \right)}{\Gamma \left ( -\frac{1}{4}-i\tilde{X}
\right)}r_{H}^{3},  \label{E7}
\end{equation}
where $\tilde{X}=X |_{\kappa \rightarrow 0}=%
\frac{\omega-\sqrt{\omega^{2}-r_{H}^{4}/4}}{2r_{H}^{2}}$ and $\tilde{Y}=Y
|_{\kappa \rightarrow 0}=\frac{\omega+\sqrt{\omega^{2}-r_{H}^{4}/4}}{%
2r_{H}^{2}}$. \bigskip

For further simplification, we need to have a closer look at the behavior of the complex gamma functions. To be able to do so, one can take advantage of a mathematical trick that can be stated as
\begin{equation}
\frac {\Gamma \left ( b+iy \right)}{\Gamma \left ( a+iy \right)}=\frac{\Gamma \left ( b+iy \right)}{\Gamma \left ( 1-a+iy \right)}\frac {\Gamma \left ( 1-a+iy
\right)}{\Gamma \left (a+iy \right)}\frac{\Gamma \left ( a-iy \right)}{\Gamma \left ( a-iy \right)},  \label{E10}
\end{equation}
where $a,b\in \mathbf{R}$. The standard complex gamma functions have the features \cite{maths,mathbook}
\begin{equation}
\Gamma \left ( 1-a+iy \right) \Gamma \left ( a-iy \right)=\frac{\pi}{sin \left [ \pi \left (
a-iy \right) \right ]},  \label{Ee}
\end{equation}
and
\begin{equation}
\left |\Gamma \left ( a+iy \right) \right | ^{2}=\Gamma \left ( a-iy \right)\Gamma \left ( a+iy \right),   
\end{equation}
which enable Eq. (\ref{E10}) to be rewritten as
\begin{equation}
\frac {\Gamma \left ( b+iy \right)}{\Gamma \left ( a+iy \right)}=\frac{1}{%
\left |\Gamma \left ( a+iy \right) \right | ^{2}}\frac {\Gamma \left ( b+iy
\right)}{\Gamma \left (1- a+iy \right)}\frac{\pi}{sin \left [ \pi \left (
a-iy \right) \right ]}.  \label{EEE}
\end{equation}
Therefore, for our case, the complex gamma function ratios of
the retarded Green's function can be expressed as
\begin{equation}
\frac {\Gamma \left ( \frac{5}{4}-i\tilde{X} \right)}{\Gamma \left ( -\frac{1%
}{4}-i\tilde{X} \right)}=\frac{1}{\left |\Gamma \left ( -\frac{1}{4}-i\tilde{%
X} \right) \right | ^{2}}\frac{\pi}{sin \left [ \pi \left ( -\frac{1}{4}+i%
\tilde{X} \right) \right ]},  \label{E8}
\end{equation}
and
\begin{equation}
\frac {\Gamma \left ( \frac{5}{4}-i\tilde{Y} \right)}{\Gamma \left ( -\frac{1%
}{4}-i\tilde{Y} \right)}=\frac{1}{\left |\Gamma \left ( -\frac{1}{4}-i\tilde{%
Y} \right) \right | ^{2}}\frac{\pi}{sin \left [ \pi \left ( -\frac{1}{4}+i%
\tilde{Y} \right) \right ]}.  \label{E9}
\end{equation}
As we are interested in the low energy limit, i.e. the regime of interest retains $ w<<r_{H}$, one can write
\begin{equation}
\begin{aligned} \tilde{X}&\sim \frac{\omega}{2r_{H}^{2}}-\frac{i}{4}, \\
\tilde{Y} &\sim \frac{\omega}{2r_{H}^{2}}+\frac{i}{4}. \end{aligned}
\label{E11}
\end{equation}
Substituting the low energy behavior of $\tilde{X}$ and $\tilde{Y}$ into Eqs. (\ref{E8}) and (\ref{E9}) together with the commonly referred relation $z\Gamma\left(z\right)=\Gamma\left(1+z\right)$, the modulus squared terms can be replaced by
\begin{equation}
\left |\Gamma \left ( -\frac{1}{4}-i\tilde{%
X} \right) \right | ^{2}=\frac{4r_{H}^{2}}{r_{H}^{4}+\omega^{2}} \left |\Gamma \left ( \frac{1}{2}-\frac{i\omega}{2r_{H}^{2}} \right) \right | ^{2},    \end{equation}
and
\begin{equation}
\left |\Gamma \left ( -\frac{1}{4}-i\tilde{%
Y} \right) \right | ^{2}= \left |\Gamma \left(-\frac{i\omega}{2r_{H}^{2}}\right) \right | ^{2}. \end{equation}
Since the complex gamma functions satisfy the relations \cite{maths}
\begin{equation}
\left |\Gamma \left (ib \right) \right | ^{2}= \frac{\pi}{b\sinh{(\pi b)}},  
\end{equation}
and 
\begin{equation}
\left |\Gamma \left ( \frac{1}{2}+ib \right) \right | ^{2}= \frac{\pi}{\cosh{(\pi b)}},   
\end{equation}
the ratios (\ref{E8}) and (\ref{E9}) take the following compact forms:
\begin{equation}
\begin{aligned} \frac {\Gamma \left ( \frac{5}{4}-i\tilde{X} \right)}{\Gamma
\left ( \frac{-1}{4}-i\tilde{X} \right)}&\sim \frac{i \left(
r_{H}^{4}+\omega^{2} \right)}{4r_{H}^{2}} \coth\left(\frac{\pi
\omega}{2r_{H}^{2}} \right),\\ \frac {\Gamma \left ( \frac{5}{4}-i\tilde{Y}
\right)}{\Gamma \left ( \frac{-1}{4}-i\tilde{Y} \right)} &\sim
-\frac{\omega}{2r_{H}^{2}} \tanh\left( \frac{\pi \omega}{2r_{H}^{2}} \right).
\end{aligned}  \label{E12}
\end{equation}
Note that for our case, $b=-\omega/2r_H^2$. Substituting the expressions (\ref{E12}) back into the two-point correlation function (\ref{E7}), we get
\begin{equation}
 G_{O_{+}}(\omega ,0)=-\frac{i \omega \left(
r_{H}^{4}+\omega^{2} \right)}{3r_{H}}.
  \label{Ef}
\end{equation}
Finally, letting $\omega \rightarrow 0$, the shear viscosity (\ref{E6}) of our concerned model is found to be
\begin{equation}
\eta=\frac{r_{H}^{3}}{3}.  \label{E13}
\end{equation}
To express the shear viscosity in terms of temperature, it is of benefit to replace the event horizon by $r_H=\sqrt{2\pi T}$ which results in
\begin{equation}
\eta=\zeta T^{3/2},    \label{shear}
\end{equation}
where $\zeta=\frac{2\sqrt{2\pi^{3}}}{3}$. From Eq. (\ref{shear}), one can perceive the effect of hyperscaling violation on the observables of the concerned (2+1)-dimensional dual model.
\begin{figure}[h]
\centering
\includegraphics[width=8cm,height=5cm]{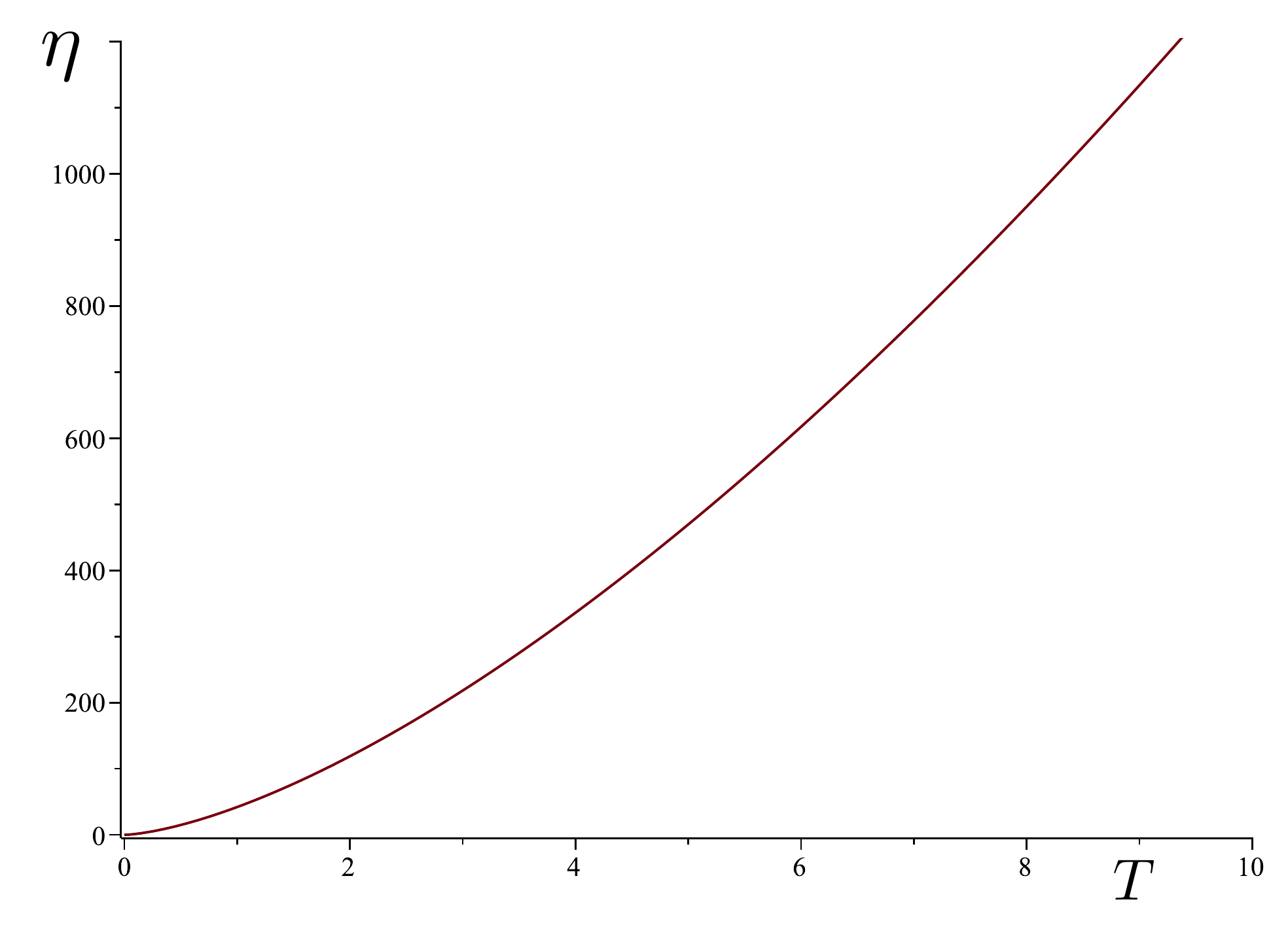}
\caption{Plot of $\protect\eta$ versus $T$ which is governed by Eq. (\ref{shear}).}
\end{figure}
\bigskip

As the \textit{hv}-factor reads $\theta=-1$, the length dimension of a spatial volume is increased by one, i.e. $d_{eff}=d_{b}-\theta$, where $d_{b}$ denotes the spatial dimension of the holographic model on the boundary. Therefore, our results indicate that for theories with anisotropy and hyperscaling violation, one can expect $\eta \propto T^{(d_{b}-\theta)/z}$ for the dual model, which is in full agreement with Refs. \cite{sh1,sh,Andrew,shh,sh2}.

\subsection{DC-Conductivity}
The DC-conductivity can be thought as the zero-frequency limit of 
\begin{equation}
\sigma^{ij}\left(\omega\right)=-\frac{1}{i\omega}\langle J^{i}\left(\omega\right) J^{j}\left(\omega\right)\rangle   \label{A}
\end{equation}
where $J^{i}$ represents the current operator \cite{Pope,Kub}. Eq. (\ref{A}) enables one to evaluate the optical conductivity of a system via Kubo's formula, and in turn, its zero-frequency limit leads to DC-conductivity which goes as
\begin{equation}
\sigma_{DC}=\lim_{\omega\to 0} \sigma^{ij}(\omega).\label{Cond}   
\end{equation}
For our case, Eq. (\ref{Cond}) reduces to
\begin{equation}
\sigma_{DC}=\frac{e^{2}}{3} \left( 2\pi \right)^{3/2} T^{3/2},  \label{E14}
\end{equation}
in which $e$ represents the charge of an electron.

\begin{figure}[h]
\centering
\includegraphics[width=8cm,height=5cm]{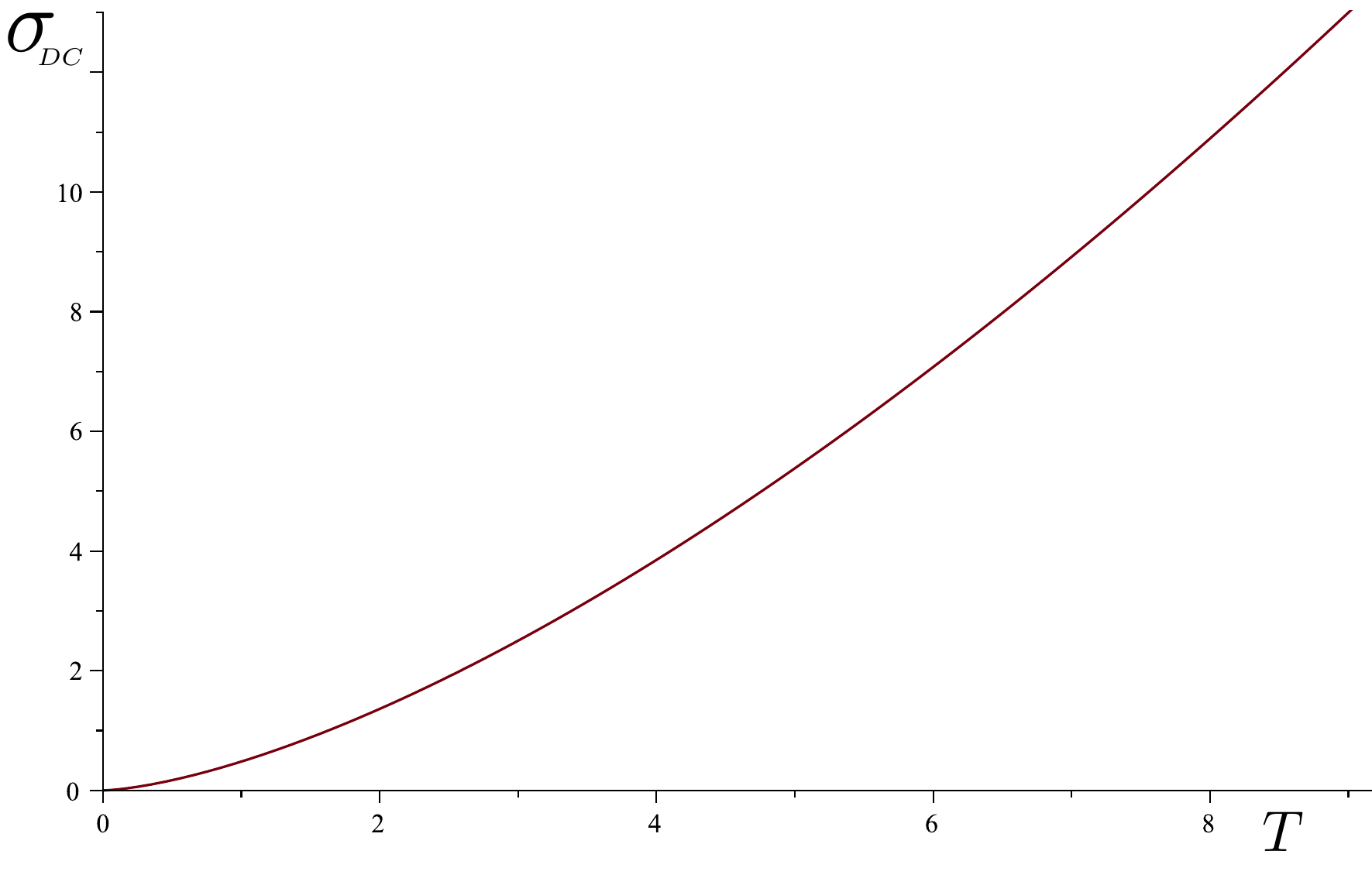}
\caption{Plot of $\protect\sigma_{DC}$ versus $T$. The plot is governed by
Eq. (\protect\ref{E14}).}
\end{figure}
\subsection{DC-Resistivity}

The DC-resistivity of the dual model for \textit{hv} Lifshitz-like BB studied throughout this paper is equivalent to the reciprocal of Eq. (\ref{E14}) which can simply be written as
\begin{equation}
\rho_{DC}=\frac{1}{\sigma_{DC}}=\frac{3}{e^{2}} \left( 2\pi \right)^{-3/2} T^{-3/2}.  \label{E15}
\end{equation}
It is worthwhile to note that the scaling behavior of the DC-resistivity carries significant information regarding the thermodynamical behavior of a system. 
\begin{figure}[h]
\centering
\includegraphics[width=8cm,height=5cm]{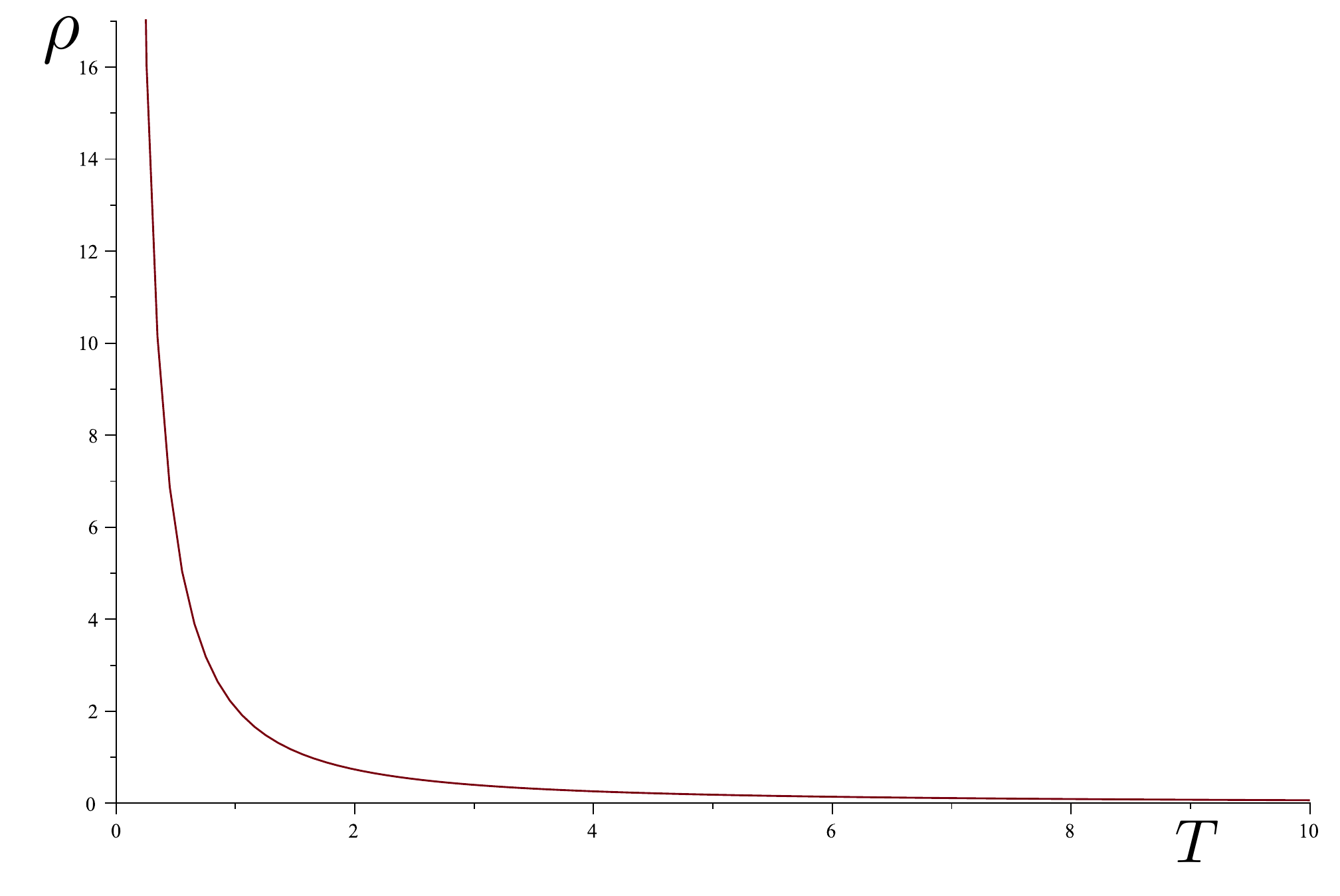}
\caption{Plot of $\protect\rho$ versus $T$. The plot is governed by Eq. (%
\protect\ref{E15}).} \label{fig3}
\end{figure}

We predict that Fig. (\ref{fig3}) represents a second order phase transition around the critical point where the graph experiences a dramatic decrease. A similar plot can be found in Ref. \cite{Pope} in which the authors first used analytical evaluations to figure out the holographic DC-conductivity of a system with arbitrary $(z,\theta)$, and subsequently discussed the behavior of the DC-resistivity for a non-relativistic system via numerical methods. Their analysis indicates that the DC-resistivity possesses different scaling behavior for different temperature regimes; namely, for $T\leq T_{critical}$ and $T> T_{critical}$. In our case, the holographic DC-resistivity behaves as $\rho_{DC} \propto T^{-3/2}$ which seems to be only a small portion of a broader picture. For studies on second order superfluid and superconducting phase transitions, one may refer to Refs. \cite{Herzog,Hartnoll}.

\section{Conclusion}
In this paper, we have computed the two-point correlation function of a non-relativistic fluid via the linear response theory and have in turn used the associated result to obtain exact expressions for the shear viscosity, DC-conductivity, and DC-resistivity of the dual model of \textit{hv} Lifshitz-like BBs with exponents $z=2$ and $\theta=-1$. The BB model of our concern exhibits Lifshitz-like symmetry and the corresponding holographic model is a member of non-relativistic CFT at strong coupling. We have obtained the two-point correlation function as $G_{O_{+}}(\omega ,0)=-i \omega \left(
r_{H}^{4}+\omega^{2} \right)/3r_{H}$, and accordingly, the transport coefficients of the dual non-relativistic model are found to exhibit the behavior $\eta \propto T^{3/2}$, $\sigma_{DC} \propto T^{3/2}$, and $\rho_{DC} \propto T^{-3/2}$ which is consistent with the general scaling behavior of models with arbitrary metric exponents; i.e.  $\eta \propto T^{(d_{b}-\theta)/z}$.\\

This study can be considered as an analytical example for applications of AdS/CFT correspondence which makes it possible to link two seemingly different phenomena occurring in different regions of spacetime to each other and treat them as a whole. Throughout our study, we have shown that it can indeed be possible for one to start from a relatively simple gravitational model and obtain some basic knowledge about the observables of a more complicated system, for instance a strongly coupled fluid, or vice versa: a variety of experimental data collected from strongly coupled systems accessible in nature can be used as a catalog for astronomical objects as long as the relevant theories are in harmony with each other.\\

In future work, we plan to extend our study by examining the relevance of our results with the theory of magnetic monopoles and D-branes from string theory and check whether it can provide us with any useful information regarding concepts like quark confinement and chiral symmetry breaking.


\begin{thebibliography}{99}

\bibitem{Dirac} P. A. M. Dirac, Proc. R. Soc. Lond. A \textbf{133}, 60 (1931).
\bibitem{Maldacena} O. Aharony, S. S. Gubser, J. Maldacena, H. Ooguri and Y. Oz, Phys. Rep. \textbf{323}, 183 (2000).
\bibitem{Witten} E. Witten, Adv. Theor. Math. Phys. \textbf{2}, 253 (1998).
\bibitem{AdS} J. M. Maldacena, Int. J. Theor. Phys. \textbf{38}, 1113 (1999).
\bibitem{Hoft} G. 't Hooft, Conf. Proc. C \textbf{930308}, 284 (1993).
\bibitem{Susskind} L. Susskind, J. Math. Phys. \textbf{36}, 6377 (1995).
\bibitem{membrane1} K. S. Thorne, R. H. Price and D.A. MacDonald, \textit{Black Holes: The Membrane Paradigm; The Silliman Memorial Lectures Series} (Yale University Press, Yale, 1986).
\bibitem{membrane2} P. Kovtun, D. T. Son, A. O. Starinets, JHEP \textbf{0310}, 064 (2003).
\bibitem{numeric1} J. Greensite, Prog. Part. Nucl. Phys. \textbf{51}, 1 (2003).
\bibitem{numeric2} K. I. Kondo, A. Shibata, T. Shinohara and S. Kato, Phys. Rev. D \textbf{83}, 114016 (2011).
\bibitem{numeric3} G. P. Engel, L. Giusti, S. Lottini and R. Sommer, Phys. Rev. Lett. \textbf{114}, 112001 (2015).
\bibitem{numeric4}  H. J. Rothe, \textit{Lattice Gauge Theories} (World Scientiﬁc, Singapore, 2012).
\bibitem{numeric5}  J. Greensite, \textit{An Introduction to the Conﬁnement Problem} (Springer, Berlin, 2011).
\bibitem{h1}K.~Copsey and R.~Mann,
JHEP \textbf{04}, 079 (2013).
\bibitem{h2}Y.~Lei and S.~F.~Ross,
Class. Quant. Grav. \textbf{31}, 035007 (2014).
\bibitem{h3}I.~Papadimitriou,
Nucl. Part. Phys. Proc. \textbf{273-275}, 1487-1493 (2016).
\bibitem{h4}A.~Karch,
JHEP \textbf{06}, 140 (2014).
\bibitem{h5}L.~Li,
Phys. Lett. B \textbf{767}, 278-284 (2017).
\bibitem{h6}J.~P.~Wu and X.~M.~Kuang,
Phys. Lett. B \textbf{753}, 34-40 (2016).
\bibitem{h7}M.~A.~Ganjali,
Phys. Rev. D \textbf{93}, no.2, 024002 (2016).
\bibitem{h8}M.~Alishahiha, E.~O Colgain and H.~Yavartanoo,
JHEP \textbf{11}, 137 (2012).
\bibitem{h9}S.~Mukhopadhyay and C.~Paul,
Nucl. Phys. B \textbf{938}, 571-593 (2019).
\bibitem{h10}J.~F.~Pedraza, W.~Sybesma and M.~R.~Visser,
Class. Quant. Grav. \textbf{36}, no.5, 054002 (2019).
\bibitem{h11}J.~F.~Pedraza, W.~Sybesma and M.~R.~Visser,
Class. Quant. Grav. \textbf{36}, no.5, 054002 (2019).
\bibitem{h12}E.~Kiritsis and Y.~Matsuo,
JHEP \textbf{12}, 076 (2015).
\bibitem{exponent}  M. Park, J. Park and J. H. Oh, Eur. Phys. J. C \textbf{77}, 810 (2017). 
\bibitem{Hooft} G. 't Hooft, Nucl. Phys. B \textbf{72}, 461 (1974).
\bibitem{Veneziano} G. Veneziano, Il Nuovo Cimento A \textbf{57}, 190 (1968).
\bibitem{dual2} H.B.Nielsen and P. Olesen, Nucl. Phys. B \textbf{61}, 45 (1973).
\bibitem{Mandelstam}  S. Mandelstam, Phys. Lett. B \textbf{53}, 476 (1975).
\bibitem{dual3}  G. 't Hooft, Nucl. Phys. B \textbf{190}, 455 (1981).
\bibitem{dual} L. D. D. A. D. Giacomo and G. Paffuti, Phys Lett. B \textbf{349}, 513 (1995).
\bibitem{NOST}P.~Orland,
Nucl. Phys. B \textbf{428}, 221-232 (1994).
\bibitem{Nambuu}Y.~Nambu,
Nucl. Phys. B \textbf{579}, 590-616 (2000).
\bibitem{z} J. A. Hertz, Phys. Rev. B \textbf{14}, 1165 (1976).
\bibitem{8} S. Kachru, X. Liu and M. Mulligan, Phys. Rev. D \textbf{78}, 106005 (2008).
\bibitem{Hooft2} G. 't Hooft, Conf. Proc. C \textbf{930308}, 284 (1993).
\bibitem{biz} H. G\"{u}rsel and I. Sakall{\i}, Eur. Phys. J. C \textbf{80}, 234 (2020).
\bibitem{SourceBecar} X. H. Feng and W. J. Geng, Phys. Lett. B \textbf{31105}, 395 (2015).
\bibitem{yeni} E. Perlmutter, JHEP \textbf{06}, 165 (2012).
\bibitem{SCH} H. Singh, JHEP \textbf{07}, 082 (2012).
\bibitem{wald}R. M. Wald, \textit{General Relativity} (University of Chicago Press, Chicago, 1984).
\bibitem{hawr}S. W. Hawking, Commun. Math. Phys. \textbf{43}, 199 (1975).
\bibitem{Becar} R. Becar, P. A. Gonzalez, and Y. Vasquez, Gen. Relat. Gravit. \textbf{49}, 26 (2017).
\bibitem{kaynak} P. Q. Jin and Y. Q. Li, Phys. Rev. B \textbf{74}, 085315 (2006).
\bibitem{Kubo} M. Ammon and J. Erdmenger, \textit{Gauge/Gravity Duality} (Cambridge University Press, Cambridge, 2015).
\bibitem{Czajka:2017bod} A.~Czajka and S.~Jeon,
Phys. Rev. C \textbf{95}, no.6, 064906 (2017).
\bibitem{Moghaddas:2017itv} M. Natsuume, \textit{AdS/CFT Duality User Guide} (Springer, Japan, 2015).
\bibitem{H.Lu} Z. Y. Fan and H. Lu, JHEP \textbf{04}, 139 (2015).
\bibitem{maths} I. S. Gradshteyn and I. M. Ryzhik, \textit{Table of Integrals, Series, and Products} (Academic Press, New York, 1980).
\bibitem{mathbook} M. Abramowitz and A. Stegun, \textit{Handbook of Mathematical Functions} (Dover Publications, New York, 1970).
\bibitem{sh1} M. Ammon, M. Kaminski and A. Karch, JHEP \textbf{11}, 028 (2012).
\bibitem{sh} X. Dong, S. Harrison, S. Kachru, G. Torroba and H. Wang, JHEP \textbf{1206}, 041 (2012).
\bibitem{Andrew} A. Lucas, S. Sachdev and K. Schalm, Phys. Rev. D \textbf{89}, 066018 (2014).
\bibitem{shh} J. Bhattacharya, S. Cremonini, and A. Sinkovics, JHEP \textbf{02}, 147 (2013). 
\bibitem{sh2} K. S. Kolekar, D. Mukherjee and K.Narayan,  Phys. Lett. B \textbf{760}, 86 (2016).
\bibitem{Pope} S. Cremonini, H. S. Liu, H. Lü, and C.N. Pope, JHEP \textbf{04}, 009 (2017).
\bibitem{Kub} C. Charmousis, B. Gouteraux, B. S. Kim, E. Kiritsis and R. Meyer, JHEP \textbf{1011}, 151 (2010).
\bibitem{Herzog} C. P. Herzog, P. K. Kovtun and D. T. Son, Phys. Rev. D \textbf{79} 066002 (2009).
\bibitem{Hartnoll} S. A. Hartnoll, C. P. Herzog and G. T. Horowitz, Phys. Rev. Lett. \textbf{101}, 031601 (2008).



\end{thebibliography}
\end{document}